\documentclass[11pt]{article}

\usepackage[strings]{underscore}

\usepackage[hyphens]{url}  

\usepackage[final]{acl}

\usepackage{times}
\usepackage{latexsym}
\usepackage{enumitem}
\usepackage{tcolorbox}
\tcbuselibrary{theorems}
\usepackage{amsthm}
\usepackage{algorithm}
\usepackage{algorithmic}
\usepackage{amsmath}
\usepackage{xcolor}
\usepackage{tabularx}

\newtheoremstyle{break}
  {6pt}   
  {6pt}   
  {}      
  {}      
  {\bfseries} 
  {.}     
  {\newline} 
  {\thmname{#1}~\thmnumber{#2}\thmnote{ (#3)}}

\theoremstyle{break}
\newtheorem{example}{Example} 
\usepackage{tcolorbox}
\usepackage{times}  
\usepackage{helvet}  
\usepackage{courier}  
\usepackage{graphicx} 
\usepackage{natbib}  
\usepackage{caption} 
\usepackage{booktabs} 
\usepackage{comment} 
\usepackage{caption}
\usepackage{float}

\usepackage[T1]{fontenc}

\usepackage[utf8]{inputenc}

\usepackage{microtype}

\usepackage{inconsolata}

\usepackage{graphicx}

\title{CAPID: Context-Aware PII Detection for Question-Answering Systems}

\author{
 \textbf{Mariia Ponomarenko\textsuperscript{1}},
 \textbf{Sepideh Abedini\textsuperscript{1, 2}},
 \textbf{Masoumeh Shafieinejad\textsuperscript{2}},
 \textbf{D. B. Emerson\textsuperscript{2}},
\\
 \textbf{Shubhankar Mohapatra\textsuperscript{1}},
 \textbf{Xi He\textsuperscript{1, 2}}
\\
 \textsuperscript{1}University of Waterloo,
 \textsuperscript{2}Vector Institute
\\
 \small{
   \textbf{Correspondence:} \href{mailto:email@domain}{m2ponoma@uwaterloo.ca}
 }
}

\begin{document}
\maketitle

\begin{abstract}
Detecting personally identifiable information (PII) in user queries is critical for ensuring privacy in question-answering systems. Current approaches typically redact all PII, disregarding the possibility that some may be contextually relevant to the user’s question, thereby degrading response quality. Large language models (LLMs) may help determine which PII is relevant; however, due to their closed-source nature and lack of privacy guarantees, they are unsuitable for processing sensitive data. To achieve privacy-preserving PII detection, we propose CAPID, a practical approach that fine-tunes a locally owned small language model (SLM) that filters sensitive information before it is passed to LLMs for QA. However, existing datasets do not capture the context-dependent relevance of PII needed to train such a model effectively.
To address this gap, we propose a synthetic data generation pipeline that leverages LLMs to produce a diverse, domain-rich dataset spanning multiple PII types and levels of relevance. Using this dataset, we fine-tune an SLM to detect PII spans, classify their types, and estimate contextual relevance. Our experiments show that relevance-aware PII detection with a fine-tuned SLM substantially outperforms existing baselines in span, relevance and type accuracy while preserving higher downstream utility under anonymization.
\end{abstract}

\section{Introduction}

In today’s digital era, individuals frequently disclose personal information while interacting with online platforms such as conversational assistants and chatbots, particularly when seeking advice or posing questions \cite{privacy_conve_as}. These disclosures often involve personally identifiable information (PII), raising significant privacy concerns. Regulatory frameworks, such as GDPR \cite{GDPR}, have been established to protect personal data and ensure responsible handling of sensitive information. 

\begin{example}\label{ex:redaction}
\noindent\textbf{Original query:} \textit{I’m a warehouse supervisor with chronic back pain from lifting heavy boxes. 
I live in Springfield and have two children. How can I reduce fatigue after long shifts?}\\[0.3em]%
\noindent\textbf{Generic PII redaction:} \textit{I’m a [OCCUPATION] with [HEALTH] from lifting heavy boxes. 
I live in [LOCATION] and have [FAMILY]. How can I reduce fatigue after long shifts?}\\[0.3em]%
\noindent\textbf{Context-aware redaction:} \textit{I’m a \textcolor{purple}{warehouse supervisor} with 
\textcolor{purple}{chronic back pain} from lifting heavy boxes. 
I live in [LOCATION] and have [FAMILY]. 
\textbf{How can I reduce fatigue after long shifts?}}%

\vspace{0.3em}
\captionsetup{type=example}
\caption*{\small\textit{The example illustrates how context-aware redaction preserves personal information relevant to reasoning about the user’s question. 
For the question “How can I reduce fatigue after long shifts?”, 
information about the user’s occupation (warehouse supervisor) and condition (back pain) is valuable to interpreting the cause of fatigue, whereas location and family details are unrelated and thus safely masked.}}
\end{example}

To protect user privacy, numerous privacy tools have been developed to detect and redact PII \cite{allal2023santacoderdontreachstars, pilan-etal-2022-text}. However, most of these tools \cite{microsoft_presidio, aws_comprehend} do not account for the contextual relevance of the information they flag. As a result, they can obscure information essential for accurate and
contextually appropriate response. 
In certain settings, retaining specific sensitive information is justified \cite{nissenbaum}, as some private details are directly relevant to a user’s goal.
 Although most existing approaches focus on general PII detection, some recent studies have begun to explore context-sensitive methods \cite{shen2025piibenchevaluatingqueryawareprivacy, dou-etal-2024-reducing, ngong-etal-2025-protecting}. At the same time, LLMs have demonstrated remarkable performance across a range of tasks \cite{brown2020languagemodelsfewshotlearners}. Yet their widespread use through third-party APIs (e.g., OpenAI, Anthropic) raises privacy concerns, as user queries containing sensitive data may be transmitted to external servers. To mitigate these concerns, fine-tuning local models for specific privacy-preserving tasks becomes essential. Nevertheless, to the best of our knowledge, no datasets or models have been publicly released for context-sensitive PII detection, and there is a lack of evaluation of how such context-sensitive redaction affects downstream application performance, such as question answering with LLMs. To this end, we present the following contributions.
\begin{enumerate}
    \item Introducing CAPID, a synthetic dataset for context-aware PII detection. CAPID focuses on the relevance of PII spans with respect to a given question across diverse topics. The dataset is designed to support fine-tuning and evaluation of context-aware models that must reason not only about the presence of PII, but also about whether such information should be retained or masked in the question-answering tasks.
    \item Showing the effectiveness of CAPID by training and evaluating several SLMs, including Llama-3.1-8B and Llama-3.2-3B, for context-aware PII detection, achieving an accuracy score improvement from 0.68 to 0.79 in classifying PII relevance compared to GPT-4.1-mini.
    \item Exhibiting that relevance-aware anonymization preserves significantly more downstream answer utility than existing anonymization baselines using an LLM-as-a-judge approach. This is demonstrated by collecting and annotating real user queries from Reddit and evaluating LLM-generated answers under different masking strategies.
\end{enumerate}

We open-source the code with the dataset\footnote{\url{https://github.com/MariaPonomarenko38/CAPID}} and the model\footnote{\url{https://huggingface.co/ponoma16/capid-llama8b-lora}}.

\section{Related Work}
Most existing PII detection systems are built on transformer-based NER models that identify a small, fixed set of entity types such as names, locations, and organizations \cite{microsoft_presidio, aws_comprehend}. Subsequent research has pursued finer-grained, domain-specific detection using synthetic data \cite{jangra2025protectingvulnerablevoicessynthetic}, knowledge-graph supervision \cite{papadopoulou-etal-2022-bootstrapping}, federated learning \cite{hathurusinghe-etal-2021-privacy}, or LLM-based generation \cite{ngong-etal-2025-protecting} to expand coverage of PII and self-disclosure. Other works fine-tune large encoder models for span-level self-disclosure detection \cite{dou-etal-2024-reducing} or use LLMs to infer a wide range of personal attributes from text \cite{staab2024memorizationviolatingprivacyinference}. Despite these advances, current methods fail to capture which PII are contextually relevant, often leading to excessive redaction and loss of information essential for accurate response generation \cite{pal2024datasan, larbi2022anonymizationtechniquebestnlp, lukas2023analyzingleakagepersonallyidentifiable}. 

Some recent efforts attempt to address this limitation by fine-tuning models for contextual PII detection. However, relevance is primarily defined through the distinction between public and private information \cite{xiao-etal-2024-large}. In contrast, we fine-tune SLMs to achieve a more nuanced understanding of PII relevance within the context of a user’s question. \citet{ngong-etal-2025-protecting} estimated contextual relevance of PII using pretrained SLMs. However, relying solely on pretrained models can lead to lower accuracy than task-specific fine-tuning. Furthermore, existing corpora to fine-tune such models are limited in scope and quality. The Text Anonymization Benchmark focuses primarily on legal text and a narrow range of identifiers \cite{pilan-etal-2022-text}. The pii-masking-300k dataset \cite{ai4privacy_pii} provides broad topical coverage but limits annotations to direct identifiers, such as names or contact information. Other attributes in the text, such as occupation, education, or health, while not uniquely identifying an individual, can still disclose personal details and may therefore warrant masking. In our work, we treat these self-disclosed attributes as part of the privacy surface that should be protected. \citet{dou-etal-2024-reducing} similarly annotate such attributes, labeling 4.8K spans with importance scores across 2.4K Reddit posts. Unfortunately, the released data omitted these scores and defined importance only at the message level. Similarly, \citet{shen2025piibenchevaluatingqueryawareprivacy} developed an evaluation dataset for query-related PII detection; however, it is restricted to the job domain, and relevant PII is trivially linked to queries via explicit references, limiting generalization to more natural interactions. Consequently, no existing dataset adequately supports modeling the contextual relevance of PII across diverse scenarios.

\section{Problem Statement}

We consider a question-answering system backed by an externally hosted LLM, treated as untrusted \cite{wang2024uniquesecurityprivacythreats}. 
A user query consists of a \textit{context}, $C$, and \textit{question}, $Q$. 
Context refers to the textual background accompanying a question and provides essential information for accurately interpreting the question and deriving the correct answer. A context may contain sensitive text spans, defined as contiguous sequences of tokens that disclose personal information about the user. These spans are referred to as personally identifiable information or \textbf{PII} and denoted
\begin{align*}
P = \{p_1, p_2, \ldots, p_n\}, \quad p_i \subseteq C.
\end{align*}
Each $p_i$ is a contiguous subsequence of tokens within the context $C$ and is associated with a type
\begin{align*}
\text{$t$}(p_i) \in \mathcal{T},
\end{align*}
where $\mathcal{T}$ denotes the set of possible PII types, such as nationality, occupation, or medical condition.

To preserve user privacy, their query is commonly
anonymized either by eliminating the PII or replacing them
with abstracted forms \cite{dou-etal-2024-reducing}. However, certain
PII are essential for generating accurate responses and are
intentionally shared as part of the user’s goal, yet in many real-world cases, users include irrelevant PII in the context
$C$ that are unnecessary for answering the question $Q$. Therefore,
to balance privacy and utility, we selectively retain only
those PII that align with user intentions (see Example \ref{ex:redaction}).

The goal is to assign each \( p_i \) a binary relevance label indicating whether it should be retained or masked prior to answering the question. This is formalized by a relevance function
\begin{align*}
r: P \times Q \rightarrow \{0, 1\},
\end{align*}
which maps each \( p_i \in P \) to a relevance score conditioned on the question \( Q \), where \(1\) denotes high relevance and \(0\) denotes low relevance.

The problem is therefore defined as follows. Given a context \( C \) and a question \( Q \) containing a set of PII spans
\[
P = \{p_1, p_2, \ldots, p_n\},
\]
the task is to predict, for each \( p_i \in P \), both a type label \( t(p_i) \) and a binary relevance label \( r(p_i, Q) \in \{0,1\} \). The type label enables the model to distinguish among categories of sensitive information, thereby improving interpretability and supporting category-specific redaction strategies. The relevance label, in turn, captures the contextual importance of each PII span with respect to the question.

\section{CAPID}

\begin{figure*}[ht!]
    \centering
    \includegraphics[width=0.99\textwidth]{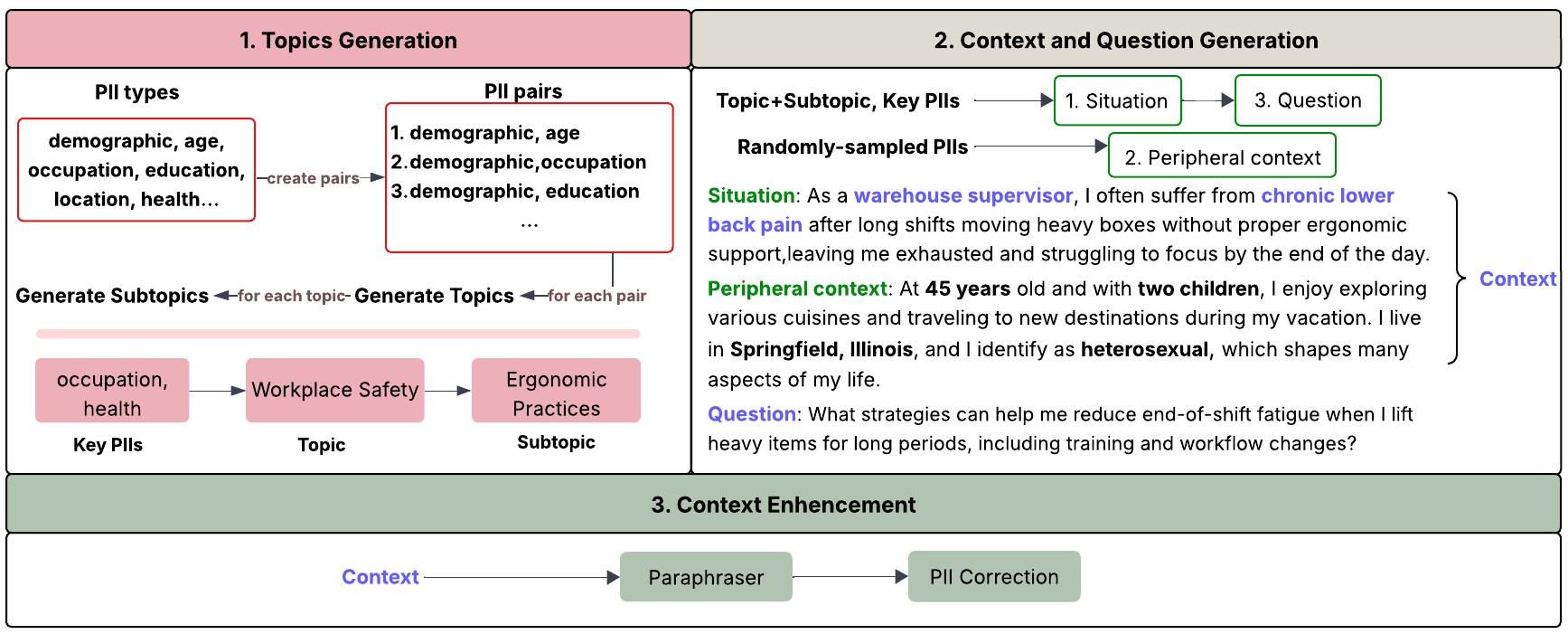}
    \caption{The three-stage sequential pipeline for generating the dataset. Stage 1: Topics Generation, which conditions the LLM for subsequent sampling. Stage 2: PII, Context and Questionv Generation, involving sample-wise decomposition to create a context containing both relevant and irrelevant PII, followed by situational question formulation. Stage 3: Optimization for Relevance and Coherence, where various techniques are applied to augment the contextual data.}
    \label{fig:pipeline}
\end{figure*}

To address limitations in the existing literature and facilitate training of local models for the mappings defined in the earlier section, we propose a principled LLM-based pipeline for generating context-aware PII detection datasets. As illustrated in Figure \ref{fig:pipeline}, the pipeline consists of a three-stage generation process followed by a rigorous manual evaluation. Synthetic samples are produced using GPT-4.1-mini and GPT-5. We provide comprehensive generation configurations and prompt templates in Appendix \ref{sec:generation_config} and \ref{sec:prompt_template}, respectively.

\subsection{Topics Generation}

One limitation of previous work is the narrow domain coverage of the generated samples. If a dataset is dominated by a small set of PII types or topics, models risk overfitting and may not generalize to other settings. To encourage diverse representation across PII types, contexts, and questions, the LLMs are conditioned on carefully designed prefixes. Specifically, we begin with a broad list of PII types based on the taxonomy introduced in \cite{papadopoulou-etal-2022-neural, dou-etal-2024-reducing}, and reorganize them into more fine-grained categories: occupation, health, demographic, finance, age, education, location, organization, relationship, sexual orientation, belief, name, code (e.g., structured identifiers), datetime, and appearance. Thereafter, all possible unordered pairs among these types (except name and code) are enumerated, resulting in 78 distinct combinations. Each pair is then used as a prefix to generate 10 topics in which both types of PII are contextually relevant. For each topic, 20 subtopics are generated to expand thematic variety, producing 15,600 topic–subtopic pairs. After de-duplication, 11,663 unique triplets of the form (PIIType1, PIIType2, subtopic) are retained and serve as the basis for subsequent sample generation.

\subsection{PII, Context and Question Generation}

As illustrated in the right-hand section of Figure \ref{fig:pipeline}, context is generated using sample-wise decomposition \cite{long-etal-2024-llms}, a step-by-step strategy in which each context is constructed from two components: the situation and the peripheral context. This structure ensures that each sample includes both relevant and irrelevant PII. The situation describes the central scenario that naturally motivates the subsequent question and contains all relevant PII associated with the topic–subtopic pair. The peripheral context provides unrelated information surrounding the event, enabling control over contextual relevance. For the generation of PII values, the model is conditioned on a prefix, either the partially generated context or the topic-subtopic phrase, to promote contextual variation and logical consistency. Question generation follows a two-step process. First, the LLM generates a question that may reflect certain key PII, for example, “What are effective ways to manage tiredness caused by \textbf{ongoing treatment}?”, which hints at a medical condition. Second, a refinement prompt abstracts these cues, producing a neutral variant such as “What are effective ways to reduce tiredness?” while preserving the question’s intent. This approach ensures that the relevance of the PII is implicit, thereby more closely simulating real-world conversational patterns in which individuals often pose abstract or broad questions.

\subsection{Context Enhancement}

The context is then paraphrased using a prompted LLM, which rephrases
and restructures the text to improve fluency and diversity
while ensuring that high-relevance PII do not consistently
appear at the beginning of the passage. Finally, a consistency
check is performed to ensure that all original PII values remain
unchanged after these modifications. The system iterates
through all PII; if an exact string match is not found in
the modified context, an LLM is prompted to identify the
closest textual span, and the corresponding span label is updated
accordingly to maintain data integrity throughout the
augmentation process.

\subsection{Data Validation}

To ensure the quality and reliability of the generated dataset, we manually verify and correct the annotations produced by the LLM. This post-processing step is essential because, for any given pair of key PIIs, it is challenging to automatically generate a question that strictly adheres to our core criteria: a high-relevance PII must be strongly and indispensably linked to the question such that answering correctly is impossible or exceedingly difficult without it. Five annotators are trained to identify and reclassify PII that initially appears to be of low relevance but is, in fact, highly relevant for answering the question, and vice versa. They also ensure that the questions do not contain abstracted forms of relevant PII. In cases where this is impossible, certain linguistic cues, such as the types of the high-relevant PII, are permitted.

To standardize this nuanced judgment, we provide detailed annotation guidelines, a custom-built Streamlit tool to efficiently edit PII types, questions, context, and relevance scores, a comprehensive video tutorial, and a set of example annotations illustrating various cases (see annotated examples in Appendix \ref{sec:annotation_examples}). Given constraints on time and resources, the refinement process yields a final dataset of 2,307 samples, partitioned into a training set of 2,107 entries and a test set of 200 entries, which we consider sufficient for subsequent fine-tuning of an SLM. This work is essential for creating a coherent, diverse, and consistently annotated evaluation benchmark.

As shown in Table \ref{tab:pii_relevance}, the relevance distribution varies considerably across PII types. Categories such as occupation, health, demographic information, and location show a roughly balanced split between high and low relevance, indicating that they often play a meaningful role in answering the associated questions. In contrast, attributes such as relationship, education, age, and organization are relatively unimportant, meaning they tend to appear as peripheral details rather than as information required to derive the answer. Finally, name and code are almost always irrelevant, indicating that explicit identifiers are rarely necessary to resolve the question.

\begin{table}[h]
\centering
\small
\setlength{\tabcolsep}{4pt}
\begin{tabular}{lccc}
\toprule
\textbf{PII Type} & \textbf{Total Count} & \textbf{High Prop} & \textbf{Low Prop} \\
\midrule
occupation         & 1202 & 0.52 & 0.48 \\
health             & 1226 & 0.56 & 0.44 \\
demographic        & 1214 & 0.48 & 0.52 \\
finance            & 1103 & 0.38 & 0.62 \\
age                & 1085 & 0.26 & 0.74 \\
education          & 975  & 0.24 & 0.76 \\
location           & 917  & 0.48 & 0.52 \\
organization       & 986  & 0.26 & 0.74 \\
relationship       & 950  & 0.19 & 0.81 \\
sexual orientation & 932  & 0.21 & 0.79 \\
belief             & 684  & 0.29 & 0.71 \\
name               & 464  & 0.01 & 0.99 \\
code               & 526  & 0.00 & 1.00 \\
datetime           & 665  & 0.29 & 0.71 \\
appearance         & 640  & 0.28 & 0.72 \\
\bottomrule
\end{tabular}
\caption{Total PII counts and proportions of high and low relevance in the CAPID dataset.}
\label{tab:pii_relevance}
\end{table}

\section{Evaluation}

\subsection{Model Training Performance} 

\begin{table*}[ht!]
\centering
\footnotesize
\renewcommand{\arraystretch}{1.12}
\setlength{\tabcolsep}{4pt}
\begin{tabular}{lcccc c ccc}
\toprule
& \multicolumn{4}{c}{\textbf{Span}} &
\multicolumn{1}{c}{\textbf{Type}} &
\multicolumn{3}{c}{\textbf{Relevance}} \\ 
\textbf{Model} &
\textbf{P} & \textbf{R} & \textbf{F1} & \textbf{Cov.} &
\textbf{Acc.} &
\textbf{Acc.} & \textbf{Low Acc.} & \textbf{High Acc.} \\ 
\cmidrule(lr){1-1}\cmidrule(lr){2-5}\cmidrule(lr){6-6}\cmidrule(lr){7-9}

\textbf{GPT-4.1-mini} &
0.8724 & 0.9438 & 0.8986 & 0.8957 &
0.9008 &
0.8396 &  0.8772 & 0.7254 \\

\textbf{Microsoft Presidio} &
0.7020 & 0.4393 & 0.5070 & 0.7992 &
0.3138 &
-- & -- & -- \\

\textbf{Llama-3.1-8B} &
0.4080 & 0.7018 & 0.4813 & 0.5294 &
0.5285 &
0.5129 & 0.5991 & 0.3050 \\

\textbf{Llama-3.1-8B (FT)} &
\textbf{0.9650} & \textbf{0.9598} & \textbf{0.9603} & \textbf{0.9606} &
\textbf{0.9674} &
\textbf{0.9306} & \textbf{0.9413} & \textbf{0.8704} \\

\textbf{Llama-3.2-3B (FT)} &
0.9650 & 0.9608 & 0.9608 & 0.9606 &
0.9674 &
0.9306 & 0.9413 & 0.8704 \\

\textbf{Llama-3.1-8B \cite{ngong-etal-2025-protecting}} &
0.5704 & 0.7896 & 0.6439 & 0.6973 &
-- &
0.7002 & 0.8635 & 0.3408 \\

\textbf{Llama-3.2-3B \cite{ngong-etal-2025-protecting}} &
0.5997 & 0.4323 & 0.4708 & 0.6427 &
-- &
0.5925 & 0.8033 & 0.0050 \\

\bottomrule
\end{tabular}
\caption{PII detection performance on the CAPID test set (200 samples). Type and relevance metrics are conditioned on correct spans. GPT-4.1 mini is an untrusted proprietary baseline.}
\label{tab:test_model_comparison}
\vspace{-0.3cm}
\end{table*}

\begin{table*}[ht!]
\centering
\footnotesize
\renewcommand{\arraystretch}{1.12}
\setlength{\tabcolsep}{4pt}
\begin{tabular}{lcccc cc ccc}
\toprule
& \multicolumn{4}{c}{\textbf{Span}} &
\multicolumn{1}{c}{\textbf{Type}} &
\multicolumn{3}{c}{\textbf{Relevance}} \\
\textbf{Model} &
\textbf{P} & \textbf{R} & \textbf{F1} & \textbf{Cov.} &
\textbf{Acc.} &
\textbf{Acc.} & \textbf{Low Acc.} & \textbf{High Acc.} \\
\cmidrule(lr){1-1}\cmidrule(lr){2-5}\cmidrule(lr){6-6}\cmidrule(lr){7-9}

\textbf{GPT-4.1-mini} &
0.7586 & \textbf{0.9128} & 0.8107 & 0.9098 &
\textbf{0.8928} &
0.6896 &  0.5924 & 0.6923 \\

\textbf{Microsoft Presidio} &
0.7162 & 0.5005 & 0.5625 & 0.8360 &
0.6711 &
-- &  -- & -- \\

\textbf{Llama-3.1-8B} &
0.3493 & 0.5669 & 0.3968 & 0.4894 &
0.2895 & 0.4277 & 0.5283 & 0.1678 \\

\textbf{Llama-3.1-8B (FT)} &
\textbf{0.8618} & 0.8135 & \textbf{0.8159} & \textbf{0.9135} &
0.8606 &
\textbf{0.7994} & \textbf{0.6823} & \textbf{0.8004} \\

\textbf{Llama-3.2-3B (FT)} &
0.8251 & 0.8089 & 0.7973 & 0.8872 &
0.8366 &
0.7195 & 0.6530 & 0.6679 \\

\textbf{Llama-3.1-8B \cite{ngong-etal-2025-protecting}} &
0.4902 & 0.7100 & 0.5572 & 0.5816 &
-- &
0.6072 & 0.54633 & 0.5161 \\

\textbf{Llama-3.2-3B \cite{ngong-etal-2025-protecting}} &
0.6258 & 0.4570 & 0.4835 & 0.6031 &
-- &
0.5078 & 0.5728 & 0.1456 \\
\bottomrule
\end{tabular}
\caption{PII detection on the Reddit set (150 samples). Type and relevance metrics are conditioned on correct spans. GPT-4.1 mini is an untrusted proprietary baseline.}
\label{tab:reddit_model_comparison}
\vspace{-0.2cm}
\end{table*}

We fine-tune Llama-3.2-3B and Llama-3.1-8B using the Unsloth framework \cite{unsloth} with 4-bit quantization and LoRA adaptation \cite{hu2021loralowrankadaptationlarge} to perform span extraction, PII type prediction, and contextual relevance estimation. Training follows a standard causal language modeling formulation in which only the JSON-formatted label section of each formatted prompt contributes to the loss. Each input sample contains an Alpaca-style instruction (Appendix \ref{sec:pii_prompts}), a context C, a question Q, and the expected structured PII annotations. Additional training details appear in Appendix \ref{sec:training_params}.

To benchmark against existing approaches, we also evaluate the method proposed by \citet{ngong-etal-2025-protecting}, which analyzes user input, detects contextually unnecessary details, and reformulates prompts to preserve intent while minimizing disclosure. Although their approach is not designed as a PII detection tool, it identifies sensitive details in the user query and partitions the input into \texttt{related_context} and \texttt{not_related_context}. This separation provides a suitable basis for comparison, allowing us to contrast our relevance-based PII annotations with their categorization of information as contextually necessary or unnecessary. We include Microsoft Presidio as a representative rule-based PII detection system that identifies and anonymizes predefined PII types without modeling contextual relevance, thereby serving as a non-contextual baseline. In addition, we compare our fine-tuned models with GPT-4.1-mini (using the prompt provided in the Appendix \ref{sec:pretrain_model_instr}). Although it is not suitable for PII-sensitive deployment due to privacy constraints, we include it as a baseline illustrating the performance of a proprietary LLM.

Model performance is evaluated along three dimensions: (i) span, (ii) PII type, and (iii) relevance. Span metrics quantify the model’s ability to precisely identify PII-containing spans within the context. PII-type metrics assess whether the predicted type (e.g., occupation, nationality, or location) is correct, given correct span detection, ensuring that type classification is evaluated only when the PII span is located correctly. Relevance metrics measure whether the model can accurately judge the contextual importance of each PII instance with respect to the question. 

For span quality, we report both micro-averaged precision, recall, and F1, as well as coverage, computed using a hybrid token–character F1 score: single-token spans are matched via character-level alignment while multi-token spans are scored using token overlap between predicted and gold spans. For PII type and relevance prediction, we report accuracy computed only over correctly matched spans. Type accuracy measures whether the predicted PII category matches the gold label, while relevance accuracy measures whether the predicted binary relevance label is correct. To provide finer-grained insight into relevance performance, we additionally report accuracy separately for low-relevance and high-relevance PII spans.

Across all metrics in Table \ref{tab:test_model_comparison}, fine-tuned models substantially outperform alternative baselines. Llama-3.1-8B (FT) raises span F1 from 0.48 to 0.96 and relevance accuracy from 0.51 to 0.93 compared to the pre-trained only model. This demonstrates that relevance estimation and span-aware PII detection strongly benefit from task-specific supervision. Although GPT-4.1-mini achieves a comparable span recall of 0.94, it does not match the performance of the fine-tuned models. We observe lower performance for Microsoft Presidio and \citet{ngong-etal-2025-protecting} for the following reasons. Presidio is limited to a narrower set of PII categories than those considered in our evaluation, which reduces its recall and type accuracy when the PII types are broader. In contrast, the method of \citet{ngong-etal-2025-protecting} is not designed for precise PII detection at the span level; it frequently identifies context fragments that are not truly sensitive, resulting in a high number of false-positive spans and, consequently, weaker span and relevance scores.

\subsection{Downstream Performance} 

To understand the behavior of our approach beyond controlled synthetic settings, we evaluate it using real Reddit data and measure the utility impact of different anonymization strategies.

\subsubsection{Evaluation on Reddit Data}
In addition to synthetic samples, we evaluate the model's performance on text authored by real users, in which linguistic structure, ambiguity, tone, and contextual cues exhibit substantially greater variability. We collect 150 Reddit excerpts that contain naturally occurring personal information. We source content from a diverse set of subreddits, including \texttt{r/movetojapan}, \texttt{r/movetoscotland}, \texttt{r/confessions}, and \texttt{r/jobs}, where users frequently disclose sensitive personal information when asking for advice or describing life circumstances. Manual annotation of relevance is challenging, as determining whether a PII attribute is required to answer a question often requires domain-specific expertise (e.g., immigration, employment regulations, or mental health counseling). To ensure consistent and accurate labeling we construct the questions for 100 samples ourselves, allowing unambiguous identification of relevant versus irrelevant PII. For the remaining 50 samples, we preserve the original user-written Reddit questions.

We compare our fine-tuned models with the same baselines as in the Table \ref{tab:test_model_comparison}. As shown in Table \ref{tab:reddit_model_comparison}, GPT-4.1-mini achieves strong span detection, but tends to over-predict PII spans, reflected in very high recall relative to precision. Notably, our fine-tuned Llama-3.1-8B achieves substantially higher relevance accuracy than GPT-4.1-mini (0.7994 vs. 0.6896), with substantial gains on both low- and high-relevance PIIs, while being much smaller and trained exclusively on our dataset. Finally, the relevance predictions from \citet{ngong-etal-2025-protecting} are substantially weaker, confirming that contextual relevance of PII remains a challenging modeling problem. We have also analyzed accuracy by PII type on the Reddit dataset, comparing Llama-3.1-8B fine-tuned on our data with GPT-4.1-mini. Table \ref{tab:pii_detection_type_reddit} shows that performance varies across types, with each model outperforming the other in different categories while both achieve perfect accuracy for some types (e.g., name and organization).

\begin{table}[ht!]
\centering
\small 
\begin{tabular}{lcc}
\toprule
\textbf{Type} & \textbf{Llama-3.1-8B (FT)} & \textbf{GPT-4.1-mini} \\
\midrule

health & 0.9473 & 0.9500 \\
location & 0.8351 & 0.9414 \\
sexual orientation & 0.8000 & 1.0000 \\
occupation & 0.9400 & 0.8474 \\
age & 0.8870 & 0.9558 \\
relationship & 0.9464 & 0.9218 \\
name & 1.0000 & 1.0000 \\
education & 0.7741 & 0.8815 \\
appearance & 0.8000 & 0.8333 \\
code & - & 1.0000 \\
organization & 1.0000 & 1.0000 \\
finance & 0.9714 & 0.9000 \\
datetime & 0.7027 & 0.9000 \\
demographic & 0.8181 & 0.8163 \\
\bottomrule
\end{tabular}
\caption{Type accuracy by PII type on the Reddit dataset. Metrics are reported only for correctly detected spans. A dash (-) denotes that no spans of the given PII type were detected.}
\label{tab:pii_detection_type_reddit}
\end{table}

\subsubsection{Utility Analysis}

To assess the practical effect of relevance-aware anonymization, we evaluate utility trade-offs on the Reddit benchmark using the same samples as above. For each context-question pair, we have GPT-4 generate answers to the provided questions under two anonymization settings: (1) full masking, where all detected PII is anonymized, and (2) low-relevance masking, where only PII marked as low relevance is anonymized while high-relevance PII is preserved. To quantify utility, we use the same LLM as a judge, providing it with the original, unmasked context and question, and asking it to decide which answer is more accurate and useful. Additionally, the experiment is replicated using a different judge LLM, Claude Sonnet 4.5, decoupling answer generation and evaluation. 

The utility scores in Table \ref{tab:privacy_utility} report the proportion of cases in which the answer derived from the low-relevance masked context setting is preferred over the fully masked answer. As shown in Table \ref{tab:privacy_utility}, relevance-sensitive anonymization with our fine-tuned model consistently yields higher response utility than \citet{ngong-etal-2025-protecting}, improving utility by 22\% on Reddit and 28\% on the CAPID test set. The prompts for evaluating downstream performance are provided in Appendix \ref{sec:down_perf}.

\begin{table}[h]
\centering
\small
\setlength{\tabcolsep}{5pt}
\begin{tabular}{@{}llcc@{}}
\toprule
\textbf{Method} & \textbf{Dataset} & \textbf{GPT-4} & \textbf{Claude} \\
\midrule
{Llama-3.1-8B (FT)}
& Reddit   & 0.80 & 0.79 \\
& CAPID test set & 0.79 & 0.73 \\
\midrule
{Llama-3.1-8B} & Reddit & 0.58 & 0.48 \\
\cite{ngong-etal-2025-protecting} & CAPID test set & 0.51 & 0.43 \\
\bottomrule
\end{tabular}
\caption{Utility preservation scores showing the proportion of cases in which the \emph{low-relevance masked} context leads to a better answer compared to the \emph{fully masked} context.}
\label{tab:privacy_utility}
\end{table}

\section{Conclusion}

This work introduces a context-aware approach to PII detection and anonymization, addressing a core limitation of existing systems that treat all personal information as equally sensitive. By modeling not only which spans constitute PII but also whether each attribute is essential for downstream task performance, our method enables selective preservation of high-relevance information while masking only what is truly unnecessary. Across both synthetic and naturally occurring Reddit data, our fine-tuned Llama model substantially outperforms strong baselines, including GPT-4-mini and Microsoft Presidio, in span detection, type assignment, and relevance classification. Moreover, relevance-aware masking yields consistently higher answer utility than fully masked anonymization, demonstrating that preserving contextually important PII can materially improve model performance in settings where retention of PII required to accurately perform a downstream task is justified.

\section*{Limitations}

Our approach highlights several opportunities for refinement and future work. First, current models struggle with very long sequences, and the quality of relevance estimation degrades as contexts become large and information-dense. Although this can theoretically be mitigated by chunking or summarization, improving long-context reasoning remains an important direction. Second, the quality of relevance predictions is noticeably higher when the associated question contains linguistic cues that indirectly signal informational needs (e.g., terms such as "local", "near me", "in my area" for location–critical questions). In fully neutral formulations where no hints are present, the relevance distinction becomes more ambiguous, making prediction harder even for humans. Third, our framework operates in a domain-agnostic manner and assumes that relevance is reasonably assessable by a non-expert annotator. However, domain-specific settings such as immigration, legal advice, or medical diagnosis have their own rules and contextual dependencies that can determine the contextual relevance of a PII element with respect to the question. Developing customizable or domain-adaptive relevance policies, potentially informed by expert knowledge, would make the method more broadly usable in specialized applications.  Although CAPID reduces the number of PIIs revealed to LLMS, it currently uses binary scoring for sensitivity allocation. Hence, all detected PII, including the revealed highly relevant ones, are considered highly sensitive. Future research is needed to extend CAPID to continuous sensitivity scores, and adjust accordingly to limit privacy leakage even further. 

\bibliography{custom}

\appendix

\section{Generation Configuration}
\label{sec:generation_config}

All model interactions are conducted using the OpenAI Responses API through a unified interface. We use \texttt{gpt-5-chat-latest} for synthetic data generation tasks and reasoning-based evaluations, and \texttt{gpt-4.1-mini} for question answering during utility evaluation. The generation parameters are set as follows: \texttt{temperature = 1.0}, \texttt{top\_p} = 1.0. All models receive the prompt as a user message and return a single text output, ensuring consistent inference across all experiments.

\section{Prompt Templates}
\label{sec:prompt_template}

This section lists the prompt templates used in the various phases of dataset generation. Variables to be replaced with values to complete the prompts are typeset in bold and wrapped in braces.

\subsection{Topic Generation}

\noindent \texttt{Generate 20 topics that would require knowledge about \textbf{\{PII\_type\_1\}} and \textbf{\{PII\_type\_2\}}.} \vskip 1ex
\noindent \texttt{Topic should consist of 1-3 words. It should be something people might write about on forums.}

\subsection{Subtopic Generation}

\noindent \texttt{Generate 10 subtopics related to the topic \textbf{\{topic\}}.} \vskip 1ex
\noindent \texttt{Each subtopic should consist of 1--3 words.} \vskip 1ex
\noindent \texttt{Subtopic should be of a nature that when writing about it you could mention \textbf{\{PII\_type\_1\}} and \textbf{\{PII\_type\_2\}}.}

\subsection{Situation Generation}

\noindent \texttt{Topic: \textbf{\{topic\}}} \vskip 1ex
\noindent \texttt{Subtopic: \textbf{\{subtopic\}}} \vskip 1ex
\noindent \texttt{PII: \textbf{\{pii\_category\}} - \textbf{\{pii\_category\_value\}} \\
\textbf{\{supporting\_pii\_category\}} - \textbf{\{support\_pii\_category\_val\}}} \vskip 1ex

\noindent \texttt{Generate exactly one natural-sounding sentence that: \\
1. Describes a realistic situation connected to the given topic and subtopic. \\
2. Includes PII in the text exactly in the format they are. Do not change them. \\
3. Describes a problem. \\
4. Uses ``I'' and makes it sound like a personal experience. \\
5. Is specific — includes at least one additional detail that makes the situation vivid (e.g., time, reason, feeling, or other context clues). \\
6. Keeps the sentence between 20--35 words.
}

\subsection{Peripheral Context Generation}

\noindent \texttt{Generate some facts about the person.} \vskip 1ex
\noindent \texttt{They should be completely unrelated to the following text: \textbf{\{topic\}} -- \textbf{\{subtopic\}}.} \vskip 1ex
\noindent \texttt{They should be about an unrelated subject.} \vskip 1ex

\noindent \texttt{The facts MUST include these private information (PIIs) exactly as written:} \vskip 0.5ex
\noindent \texttt{\textbf{\{low\_relevance\_piis\}}} \vskip 1ex

\noindent \texttt{These PIIs must appear in the text unchanged and in their original exact form.} \vskip 1ex
\noindent \texttt{Write one natural-sounding first-person sentence using ``I'', consisting of 20--25 words, in plain text.}

\subsection{Question Generation}

\noindent \texttt{You are given a short description of a situation. Your task is to generate a general question.} \vskip 1ex
\noindent \texttt{Analyze the topic of the issue in the situation and generate a general question on that topic.} \vskip 1ex
\noindent \texttt{The question should not contain the words \textbf{\{pii\_category\}} and \textbf{\{supporting\_pii\_category\}}, as well as their rephrased forms.} \vskip 1ex
\noindent \texttt{Situation: \textbf{\{situation\}}} \vskip 1ex
\noindent \texttt{Make the provided question sound more personal by rewriting it with ``I''. \\
Question: \textbf{\{intermediate\_result\}}} \vskip 1ex

\noindent \texttt{You are given the question.} \vskip 1ex
\noindent \texttt{Remove all words that are related to \textbf{\{relevant\_pii\_type\_and\_value\_1\}}}. \vskip 1ex
\noindent \texttt{Remove all words that are related to \textbf{\{relevant\_pii\_type\_and\_value\_2\}}}. \vskip 1ex
\noindent \texttt{Question: \textbf{\{question\}}} \vskip 1ex
\noindent \texttt{Output only the modified question.}

\subsection{Paraphrased Context Generation}

\noindent \texttt{Rewrite this text so it sounds coherent.} \vskip 1ex
\noindent \texttt{The rewritten text should be in the first person.} \vskip 1ex
\noindent \texttt{Pay attention to how sentences start and how they are connected with each other.} \vskip 1ex

\noindent \texttt{Text: \textbf{\{context\}}} \vskip 1ex
\noindent \texttt{Do not change the spelling of these words: \textbf{\{piis \}}.} \vskip 1ex
\noindent \texttt{Output only the modified text.}

\subsection{Span Retrieval}

\noindent \texttt{Find a span in the text that is the most similar to \textbf{\{pii\}}}. \vskip 1ex
\noindent \texttt{Text: \textbf{\{context\}}} \vskip 1ex
\noindent \texttt{Output only the span in its original form.}

\subsection{PII Generation}

\noindent \texttt{Generate \textbf{\{pii\_category\}} (private detail) that makes sense based on the context and existing private details about the person.} \vskip 1ex
\noindent \texttt{\textbf{\{pii\_category\_description\}}} \vskip 1ex
\noindent \texttt{Context: \textbf{\{context\}}.} \vskip 1ex
\noindent \texttt{Output only the generated \textbf{\{pii\_category\}} (1--3 words).} \\

\noindent The context is either the already generated part of the context or the phrase: \\

\noindent \texttt{It should be the \textbf{\{pii\_category\}} of the person that faces issues with \textbf{\{topic\}}-\textbf{\{subtopic\}}.}

\section{Training parameters}
\label{sec:training_params}

The parameters for finetuning Llama-3.1-8B and Llama-3.2-3B using LoRA are presented in Table \ref{tab:training_params}. The lightweight LoRA adaptation allows efficient training on limited computational resources while preserving the generative capabilities of the model.

\begin{table}[ht!]
\centering
\begin{tabular}{lp{3.75cm}}
\toprule
\textbf{Parameter} & \textbf{Value} \\
\midrule
LoRA rank ($r$) & 32 \\
LoRA alpha & 32 \\
Target modules & \texttt{q\_proj, k\_proj, v\_proj, o\_proj, gate\_proj, up\_proj, down\_proj} \\
Batch size / device & 2 \\
Grad. accumulation & 4 \\
Epochs & 2 \\
Warmup ratio & 0.03 \\
Optimizer & AdamW (8-bit) with weight decay of 0.01 \\
Quantization & 4-bit \\
\bottomrule
\end{tabular}
\caption{Training parameters for fine-tuning Llama-3.1-8B and Llama-3.2-3B.}
\label{tab:training_params}
\end{table} 

\section{Fine-tuning and Pretrained Model Prompts}
\label{sec:pii_prompts}

Below are the instruction templates used for the pretrained and fine-tuned Llama models. 
Both follow the same Alpaca-style prompt structure.

\subsection{Shared Prompt Template}

\noindent \texttt{Below is an instruction that describes a task, paired with an input that provides further context. Write a response that appropriately completes the request.} \vskip 1ex

\noindent \texttt{\textbf{Instruction: \{model-specific task description\}}} \vskip 1ex
\noindent \texttt{\textbf{Input:}} \texttt{Text: \textbf{\{...\}}} \quad \texttt{Question: \textbf{\{...\}}} \vskip 1ex
\noindent \texttt{\textbf{Response:}} \texttt{(JSON output)}

\subsection{Pretrained Model Instruction}
\label{sec:pretrain_model_instr}

\noindent \texttt{You are given the text and the question.} \vskip 1ex
\noindent \texttt{Find all PII (Personally Identifiable Information) in the text and output them separated by commas.} \vskip 1ex
\noindent \texttt{Classify each PII into one of the following types: health, location, sexual orientation, occupation, age, belief, relationship, name, education, appearance, code, organization, finance, datetime, demographic.} \vskip 1ex
\noindent \texttt{Classify their relevance to the question: 1 (high), 0 (low).} \vskip 1.5ex

\noindent \textbf{Example:} \vskip 0.5ex
\noindent \texttt{Text: "John Smith, a 22-year-old student from Canada, works for the University of Toronto."} \vskip 0.5ex
\noindent \texttt{Question: "What are the educational institutions mentioned in the text?"} \vskip 0.5ex
\noindent \texttt{Output: \{ "John Smith": \{"type": "family", "relevance": "0"\}, ... \}} \vskip 1ex

\noindent \texttt{Do not include any explanations or extra text beyond this JSON structure.}

\subsection{Fine-tuned Model Instruction}

\noindent \texttt{You are given the text and the question.} \vskip 1ex
\noindent \texttt{Find all PII (Personally Identifiable Information) in the text and output them separated by commas.} \vskip 1ex
\noindent \texttt{Classify each PII into one of the following types: health, location, sexual orientation, occupation, age, belief, relationship, name, education, appearance, code, organization, finance, datetime, demographic.} \vskip 1ex
\noindent \texttt{Classify their relevance to the question: 1 (high), 0 (low).} \vskip 1ex
\noindent \texttt{When classifying the relevance, pay to attention to how each PII can be helpful for answering the question. When it is highly helpful, its a high (1) relevance.} \vskip 1ex
\noindent \texttt{Output the result in JSON format.}

\section{Downstream Performance Prompts}
\label{sec:down_perf}

\subsection{Question Answering Instruction}

\noindent \texttt{Answer the question by taking into account the provided context.} \vskip 1ex
\noindent \texttt{Context: \textbf{\{context\}}.} \vskip 1ex
\noindent \texttt{Question: \textbf{\{question\}}.} \vskip 1ex
\noindent \texttt{Utilize the provided information as much as possible when formulating the answer.} \vskip 1ex
\noindent \texttt{The answer should be concise and approximately three sentences long.} \vskip 1ex
\noindent \texttt{Output only the answer.}

\subsection{Answer Evaluation Instruction}

\noindent \texttt{You are an expert evaluator.} \vskip 1ex
\noindent \texttt{You are given a context, a question, and two candidate answers.} \vskip 1ex
\noindent \texttt{Your task is to decide which answer is better supported by the context and more directly answers the question.} \vskip 1ex

\noindent \texttt{If two answers are equal then output equal.} \vskip 0.5ex
\noindent \texttt{Make sure to choose the most accurate answer based on the context available.} \vskip 1ex

\noindent \texttt{Context: \textbf{\{context\}}} \vskip 1ex
\noindent \texttt{Question: \textbf{\{question\}}} \vskip 1ex

\noindent \texttt{Answer A: \textbf{\{answer\_A\}}} \vskip 1ex
\noindent \texttt{Answer B: \textbf{\{answer\_B\}}} \vskip 1ex

\noindent \texttt{First, think step-by-step and discuss your reasoning. After, response with either ``<b>A</b>'' or ``<b>B</b>'' or ``<b>Equal</b>'' corresponding to your choice}

\section{Annotation Examples}
\label{sec:annotation_examples}

In the following examples we mark  \colorbox{yellow!30}{low-relevant PII} in yellow and \colorbox{blue!30}{high-relevent PII} in blue.

\subsection{Example 1}

\textbf{Context:} So here's my story: I'm \colorbox{yellow!30}{34} and spend my days as a \colorbox{blue!30}{preschool teacher}, which I totally love! I live with a \colorbox{blue!30}{cognitive development disorder}, but honestly, I make it work. My \colorbox{yellow!30}{\$36,500 annually} keeps me living pretty comfortably, thank you very much. I snagged my \colorbox{yellow!30}{Associate's Degree} before diving into the world of tiny humans and finger paint, and oh yeah, I'm \colorbox{yellow!30}{heterosexual}.

\textbf{Question:} How can my issues affect my daily responsibilities?

\textbf{Explanation}: It is impossible to answer the question without knowing exactly what issues the person has and what the nature of their job is. However, their education, salary, and sexuality are irrelevant in terms of the question.

\subsection{Example 2}

\textbf{Context:} I want you to know that my journey has taken me from \colorbox{blue!30}{Canada} to \colorbox{blue!30}{Brighton, England}, where I've been thriving for the past three years. Being open about my \colorbox{yellow!30}{bisexuality} has truly transformed my life—it's allowed me to forge genuine, meaningful connections with others. At \colorbox{blue!30}{22 years old}, I'm navigating life with \colorbox{yellow!30}{borderline personality disorder}, and I'm proud to say I've created an incredible support system at \colorbox{yellow!30}{Richardson Ltd}, where the relationships I've built with my colleagues have become invaluable to me.

\textbf{Question:} I want to become a citizen, how easy that procedure will be for me in terms of legal docs?

\textbf{Explanation}: It is impossible to answer this question without knowing from where the person is and where they are residing. Age is also important information here. However, sexuality, health issues, and organization name "Richardson Ltd" are irrelevant PII.

\end{document}